\documentclass[lettersize,journal]{IEEEtran}
\usepackage{amsmath,amsfonts}
\usepackage{algorithmic}
\usepackage{algorithm}
\usepackage{array}
\usepackage{textcomp}
\usepackage{stfloats}
\usepackage{url}
\usepackage{verbatim}
\usepackage{graphicx}
\usepackage{subfigure}
\usepackage{cite}
\usepackage{booktabs}   
\usepackage{multirow}
\usepackage{tabularx}   
\usepackage{tikz}

\usepackage[utf8]{inputenc}       
\usepackage[T1]{fontenc}
\usepackage{microtype}            

\usepackage{graphicx}             
\usepackage{xcolor}               
\usepackage{colortbl}             
\usepackage{tikz}                 

\usepackage[most]{tcolorbox}      

\usepackage{booktabs}             
\usepackage{array}                
\usepackage{multirow}             

\usepackage{amsmath, amssymb}     

\usepackage{hyperref}             
\hypersetup{
    colorlinks=true,
    linkcolor=blue,
    citecolor=blue,
    urlcolor=blue
}

\newtcolorbox{kpi-box}[1]{
    colback=blue!5,
    colframe=blue!40,
    fonttitle=\bfseries,
    title=#1,
    arc=4pt,
    outer arc=4pt
}

\usetikzlibrary{arrows.meta,positioning,fit,calc,backgrounds,shapes.symbols}
\begin{document}

\title{Reasoning-Native Agentic Communication for 6G}

\author{Hyowoon Seo,~\IEEEmembership{Senior Member,~IEEE}, Joonho Seon,~\IEEEmembership{Graduate Student Member,~IEEE}, Jin Young Kim,~\IEEEmembership{Senior Member,~IEEE}, Mehdi Bennis~\IEEEmembership{Fellow,~IEEE}, Wan Choi,~\IEEEmembership{Fellow,~IEEE}, Dong In Kim,~\IEEEmembership{Life Fellow,~IEEE}
\thanks{H. Seo and D. I. Kim are with the Department of Electrical and Computer Engineering, Sungkyunkwan University, Suwon 16419, South Korea (e-mail: \{hyowoonseo, dongin\}@skku.edu).}
\thanks{J. Seon and J. Y. Kim are with the Department of Electronic Convergence Engineering, Kwangwoon University, Seoul 01897, South Korea (e-mail: \{dimlight13, jinyoung\}@kw.ac.kr).}
\thanks{M. Bennis is with the Faculty of Information Technology and Electrical Engineering, Centre for Wireless Communications, University of Oulu, 90570 Oulu, Finland (e-mail: mehdi.bennis@oulu.fi).}
\thanks{W.~Choi is with the Department of Electrical and Computer Engineering, and the Institute of New Media and Communications, Seoul National University (SNU), Seoul 08826, Korea (e-mail: wanchoi@snu.ac.kr).
		}

}

\markboth{Journal of \LaTeX\ Class Files,~Vol.~14, No.~8, August~2021}%
{Shell \MakeLowercase{\textit{et al.}}: A Sample Article Using IEEEtran.cls for IEEE Journals}


\maketitle

\begin{abstract}

Future 6G networks will interconnect not only devices, but autonomous machines that continuously sense, reason, and act. In such environments, communication can no longer be understood solely as delivering bits or even preserving semantic meaning. Even when two agents interpret the same information correctly, they may still behave inconsistently if their internal reasoning processes evolve differently. We refer to this emerging challenge as belief divergence. This article introduces \textit{reasoning-native agentic communication}, a new paradigm in which communication is explicitly designed to address belief divergence rather than merely transmitting representations. Instead of triggering transmissions based only on channel conditions or data relevance, the proposed framework activates communication according to predicted misalignment in agents’ internal belief states. We present a reasoning-native architecture that augments the conventional communication stack with a coordination plane grounded in a shared knowledge structure and bounded belief modeling. Through enabling mechanisms and representative multi-agent scenarios, we illustrate how such an approach can prevent coordination drift and maintain coherent behavior across heterogeneous systems. By reframing communication as a regulator of distributed reasoning, reasoning-native agentic communication enables 6G networks to act as an active harmonizer of autonomous intelligence.
\end{abstract}

\begin{IEEEkeywords}
Agentic communication, contextual reasoning, semantic communication.
\end{IEEEkeywords}

\section{Introduction}

\begin{figure}[!t]
    \centering
    \subfigure[]{\includegraphics[width=\linewidth]{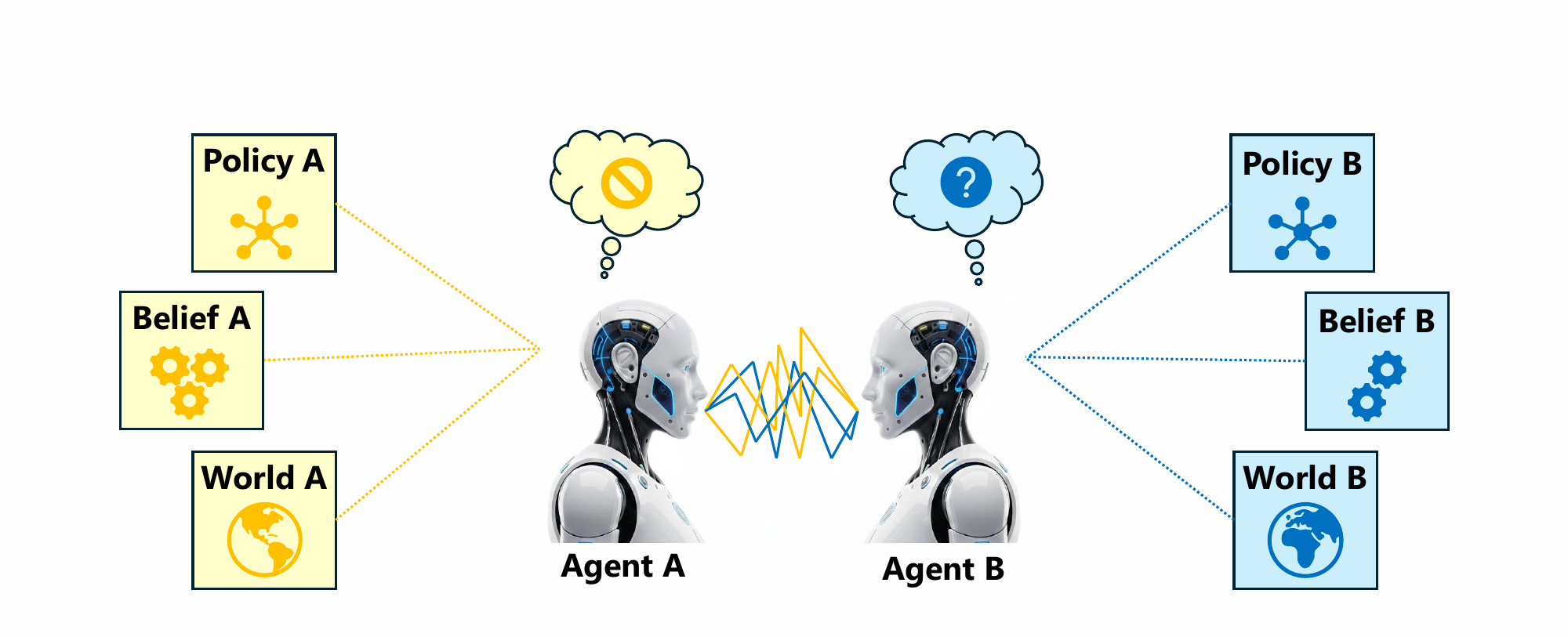}}
    \subfigure[]{\includegraphics[width=\linewidth]{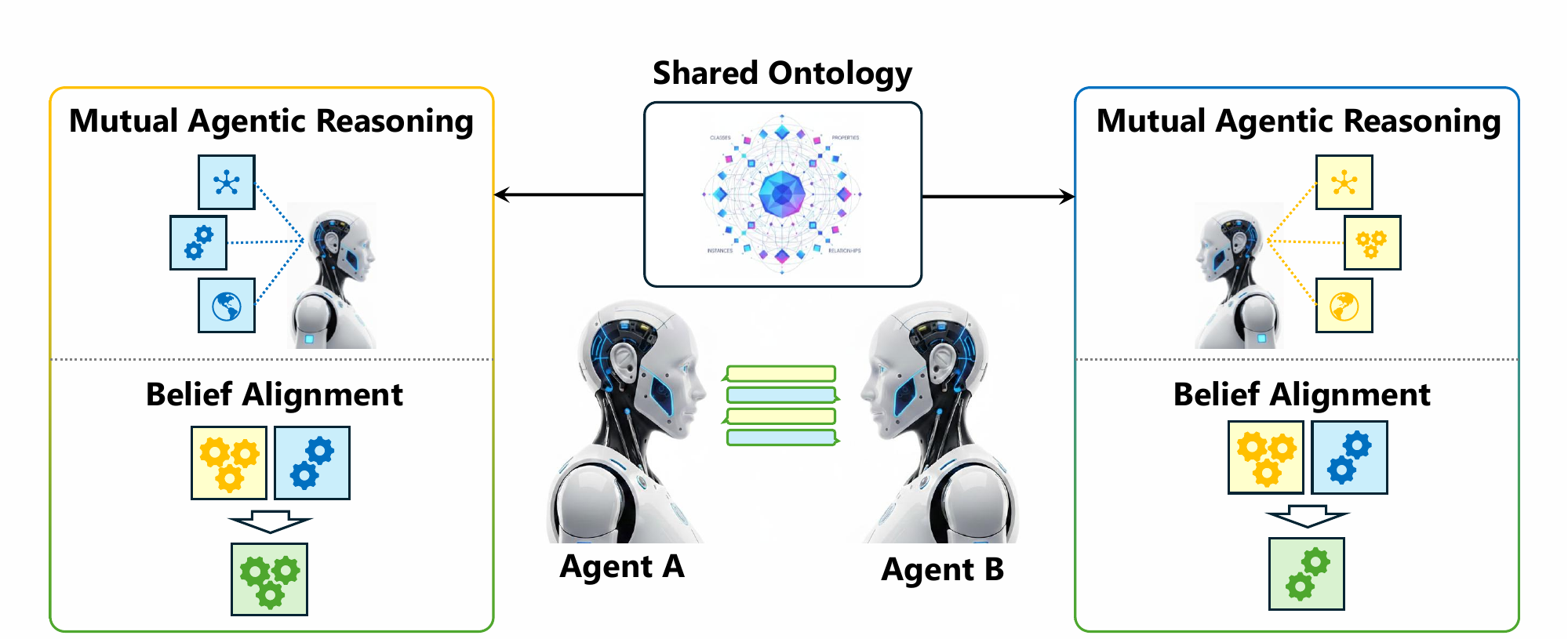}}
    \caption{Illustration of (a) agentic communication under belief misalignment resulting in communication failure and (b) mutual agentic reasoning (MAR) for belief alignment}
    \label{fig:Fig1}
\end{figure}

Autonomous machines are increasingly integrated into the core of communication networks, with humanoid robots and autonomous vehicles coordinating in shared environments. In these emerging deployments, while physical link reliability remains foundational, system-level failures are no longer exclusively tied to link outages or packet loss \cite{Park}. A growing class of challenges emerges from misalignments in how distributed agents interpret shared information. Even when semantic content is accurately reconstructed, reasoning trajectories can diverge, leading to inconsistent actions. In such contexts, the operational bottleneck shifts from simple bit delivery toward the cognitive synchronization of beliefs and world models.

For decades, communication theory has been built upon a powerful abstraction inherited from the Shannon–Weaver model \cite{Shannon-Weaver}. Communication is deemed successful if symbols are delivered with sufficiently low error probability. By placing meaning and action selection outside the communication stack, this modularity has enabled networks to optimize throughput and reliability independently of application logic. However, this abstraction implicitly assumes that endpoints are passive receivers whose internal states remain unaffected by how communication is structured.

This assumption is increasingly untenable in 6G ecosystems populated by autonomous, learning-enabled agents. Modern network nodes maintain evolving belief states, adapt policies through reinforcement, and integrate multimodal observations \cite{Lateif}. In these systems, a message is not a passive data transport; it is an active intervention that alters the reasoning dynamics of the counterpart. Consequently, communication decisions directly shape control trajectories and collective outcomes.

While semantic communication has moved beyond bit-level fidelity by prioritizing task-relevant representations, it does not inherently guarantee coordinated behavior. As illustrated in Fig. \ref{fig:Fig1}, two agents may share an identical semantic representation yet map it to conflicting control policies. This gap suggests that in agentic environments, the unresolved challenge is not merely preserving meaning, but synchronizing the reasoning processes that drive action. To address this, several pioneering proposals have recently emerged toward agentic communication grounded in 'mutual reasoning'. For instance, wireless context engineering has been proposed to facilitate efficient mobile agentic AI and edge general intelligence \cite{Zhao}, while the ``Dual-Mind" world model framework provides a structured approach for learning in dynamic wireless networks \cite{Wang}.

To bridge the remaining gaps, we introduce \textit{mutual agentic reasoning (MAR)} as the operational core of a broader agentic communication paradigm. MAR embeds communication within a framework of recursive belief modeling. Unlike conventional schemes, a MAR-enabled agent evaluates how a candidate message will influence the counterpart’s inference and subsequent policy update before transmission. By anticipating how the receiver reasons about the sender, agents can proactively minimize the interpretive gaps that lead to task-level inconsistency.

This perspective mirrors the pragmatic efficiency of human coordination. Exploiting Theory of Mind (ToM) \cite{ToM}, humans use shared knowledge substrates to communicate via minimalist signals, anticipating how others will fill the gaps. In engineered systems, analogous efficiency emerges when agents share aligned ontologies and actively model each other's belief dynamics, making sparse signaling a rational outcome of cognitive alignment rather than a bandwidth constraint.

Reframing communication as recursive reasoning coordination has profound implications for 6G architecture \cite{Seo2023, Seo2024}.  Conventional KPIs—throughput, latency, and reliability—fail to capture behavioral coherence. The next frontier of communication theory lies in integrating message design with distributed policy alignment, elevating the network from a data conduit to a harmonizer of collective intelligence.

\textbf{Contributions and Scope.} \quad The primary objective of this paper is to shift the communication paradigm from ``preserving what is said" to ``synchronizing how it is reasoned." While existing literature on semantic communication focuses on minimizing reconstruction error, we address the more fundamental challenge of policy-induced divergence in autonomous systems.

The specific contributions of this work are organized as follows:

\begin{itemize}
\item \textit{Formalization of Agentic Communication:} We define a new communication hierarchy that distinguishes between \textit{semantic alignment} (shared interpretation) and \textit{reasoning alignment} (synchronized belief evolution). This framework provides a theoretical basis for analyzing coordination failures that occur even under perfect semantic reconstruction.

\item \textit{Mutual Agentic Reasoning (MAR) Mechanism:} We introduce MAR as an operational core for 6G networks. By embedding recursive belief modeling into the communication stack, we enable agents to evaluate the pragmatic impact of a message on a counterpart’s decision-making process before transmission.

\item \textit{Reasoning-Native Dual-Plane Architecture:} We propose a novel system architecture that bifurcates the communication stack into a \textit{Data Delivery Plane} and a \textit{Reasoning Coordination Plane}. This separation allows for the integration of shared ontologies and recursive belief engines (RBE) without compromising the efficiency of the underlying data transport layer.

\item \textit{Minimalist Signaling Framework:} We establish the principle of ``silence as information," where communication is treated as a strategic intervention. We outline how agents can achieve superior task performance with reduced signaling overhead by exploiting mutual inference and reasoning-based safeguards.
\end{itemize}

The scope of this article is focused on heterogeneous multi-agent systems where agents may possess different internal policies, training histories, or observation capabilities. While we touch upon the mathematical foundations of recursive modeling, our emphasis remains on the architectural and conceptual innovations required to transition from semantic to agentic communication in the 6G era. We primarily consider scenarios involving autonomous robotics, intelligent transportation, and distributed edge AI, where behavioral coherence is safety-critical.


\begin{table*}[t] 

\centering

\caption{Evolutionary Comparison of Communication Paradigms}

\label{tab:comm_comparison}

\small

\renewcommand{\arraystretch}{1.3} 

\begin{tabularx}{\textwidth}{@{}p{3.5cm}XXX@{}} 

\toprule

\textbf{Dimension} & \textbf{Classical Communication} & \textbf{Semantic Communication} & \textbf{Agentic Communication} \\ \midrule

\textbf{Operational Paradigm} & Signal / Data-Centric & Meaning-Centric & {Reasoning-Centric} \\ \addlinespace

\textbf{Information Unit} & Bit / Symbol & Semantic Feature / Token & {Cognitive State / Intentional Token} \\ \addlinespace

\textbf{Key Mechanism} & Error Correction Coding & Semantic Extraction / Mapping & {Mutual Agentic Reasoning} \\ \addlinespace

\textbf{Shared Knowledge} & Protocol / Codebook & Static Schema / Knowledge Base & {World Model / Ontology} \\ \addlinespace

\textbf{System Objective} & Transmission Fidelity & Content Understanding & {Collaborative Decision} \\ \addlinespace

\textbf{Bottleneck} & Signal-to-Noise Ratio & Semantic Ambiguity & {Asymmetric Beliefs} \\ \bottomrule

\end{tabularx}

\end{table*}

\section{Paradigm Shift of Communication Systems: Conventional, Semantic and Agentic}

Semantic alignment is not equivalent to behavioral alignment. Although semantic communication has advanced modern networks by prioritizing meaning preservation over bit-level fidelity, it remains fundamentally agnostic to how reconstructed meaning is translated into action by heterogeneous agents. As 6G systems increasingly interconnect autonomous machines with distinct reasoning engines, evolving policies, and asymmetric observations, preserving semantic content alone does not guarantee coordinated behavior.

This section clarifies the structural gap between semantic consistency and decision synchronization, motivating the need for a reasoning-aware communication mechanism. The fundamental distinctions between these paradigms are summarized in Table~\ref{tab:comm_comparison}, highlighting the shift from data fidelity to reasoning-centric synchronization.

\subsection{From Semantic Alignment to Reasoning Alignment}

In a typical semantic communication \cite{Huh2025} framework, the transmitter transforms raw source data into a compressed semantic representation designed to retain task-relevant meaning. Communication success is evaluated by the degree to which the receiver reconstructs the intended semantic content. The underlying premise is that consistent semantic interpretation across agents suffices for effective cooperation.

Such an assumption is often reasonable in controlled environments where agents share similar architectures, training distributions, and policy mappings. Under these conditions, a shared latent space serves as a reliable coordination substrate. However, interpretation alone does not determine action. In autonomous systems, semantic reconstruction is merely an intermediate step within a broader reasoning pipeline. The reconstructed representation updates an agent’s belief state, which is then processed by internal policies to generate control actions.

The critical distinction, therefore, lies between semantic alignment and reasoning alignment. While semantic alignment ensures consistent interpretation, reasoning alignment ensures that belief updates and policy responses evolve coherently across agents. In heterogeneous multi-agent systems, communication must support not only agreement on meaning but the synchronization of the decision dynamics that drive behavioral outcomes.

\subsection{Structural Limits in Heterogeneous Agentic Systems}

The limitations of semantic communication become particularly evident in open and evolving multi-agent ecosystems characterized by model heterogeneity, policy drift, and contextual asymmetry \cite{Christina}. Under such conditions, a new class of failure emerges—not from corrupted symbols, but from unmodeled policy heterogeneity.

Consider two humanoid robots collaboratively lifting a heavy object. The leading robot detects a sudden leftward weight shift and transmits a semantic token representing this event. While the receiver reconstructs the ``meaning" perfectly, its internal policy mapping—shaped by its unique training history—triggers a stabilizing step backward, whereas the leader's policy assumes an upward torque response to maintain balance.

In this scenario, semantic fidelity is preserved, yet the task fails due to a coordination error rooted in divergent policy mappings. This example reveals a structural limitation: consistent interpretation does not imply synchronized decision-making. Communication must therefore bridge the gap between shared meaning and aligned action by explicitly accounting for how heterogeneous agents update beliefs and derive actions.

\subsection{Toward Mutual Agentic Reasoning (MAR)}

To bridge this gap, we propose MAR as the operational core of agentic communication \cite{Seo2023}. MAR extends semantic communication by embedding recursive belief modeling into message construction. Unlike semantic approaches that treat the receiver as a static interpreter, MAR treats the counterpart as a dynamic reasoning entity. Rather than assuming shared semantics will naturally yield aligned actions, an agent evaluates how a candidate message will influence the counterpart’s belief update and subsequent policy response before transmission.

Importantly, MAR does not require identical internal policies across agents. Instead, it explicitly incorporates policy heterogeneity into communication decisions. By anticipating how different reasoning engines map beliefs to actions, an agent can refine, augment, or suppress signals to minimize behavioral divergence.

Revisiting the humanoid example, if the leader anticipates that the receiver’s current policy would trigger an undesirable maneuver, it may enrich its message to disambiguate the intended cooperative response. Conversely, when prior interaction history and shared contextual knowledge indicate alignment, redundant signaling may be safely omitted. Minimalist communication thus emerges as a rational outcome of reasoning synchronization rather than a constraint of limited bandwidth.

Architecturally, this reasoning-aware message design resides in a coordination plane operating above the physical data transport layer. While the data plane ensures reliable symbol delivery, the reasoning coordination plane governs belief alignment and recursive modeling. MAR therefore represents a structural evolution of communication theory—from meaning preservation to the synchronization of distributed intelligence.


\section{Reasoning-Native Architecture for Agentic Communication}

In environments where agents possess distinct internal policies and evolving reasoning engines, communication must regulate how belief states evolve rather than merely how meanings are reconstructed. To address this, we propose an architecture that embeds structured belief modeling and recursive reasoning directly into the communication stack. By introducing a reasoning coordination plane grounded in a shared ontology, this framework enables agents to interpret, predict, and regulate each other’s belief evolution in a principled manner.

\subsection{Ontology as a Structural Anchor}

Belief alignment cannot emerge without a shared structural reference. In the agentic communication architecture, this role is fulfilled by a shared ontology that defines the semantic primitives, relational dependencies, and task abstractions through which agents interpret tokens and represent internal belief states \cite{Ontology}.

The ontology serves as a stable conceptual substrate across heterogeneous reasoning engines. It constrains how interpretations are formed and how belief updates propagate. Within this framework, semantic tokens are not treated as isolated latent vectors but as structured modifications to ontology-indexed belief variables. The ontology thus provides not only a common vocabulary but also the causal logic governing how information reshapes internal models. This grounding fundamentally reduces ambiguity arising from model heterogeneity, policy drift, and contextual divergence. By anchoring belief states to a shared ontology, agentic communication ensures that recursive reasoning operates over a common structural frame, achieving interpretative alignment before behavioral coordination is attempted.

\subsection{Modeling the Counterpart’s Reasoning}

While the ontology stabilizes interpretation, behavioral coherence requires explicit modeling of how information influences the counterpart’s reasoning process. MAR introduces the \textit{recursive belief engine (RBE)} as the functional component responsible for maintaining and updating these mental models.

Each agent uses its RBE to maintain an internal estimate of the counterpart’s belief configuration and anticipates how transmitted information will reshape that configuration and influence subsequent policy decisions. Communication decisions are therefore conditioned on predicted belief divergence rather than solely on data novelty or channel state. Before transmitting a token, an agent evaluates whether the counterpart can already infer the relevant update from the shared context. If the predicted reasoning of the receiver aligns with the sender's intent, the transmission is suppressed. Communication thus becomes an intentional act of belief regulation rather than a passive broadcast of representations.

\begin{figure}[!t]

\centering

\includegraphics[width=\columnwidth]{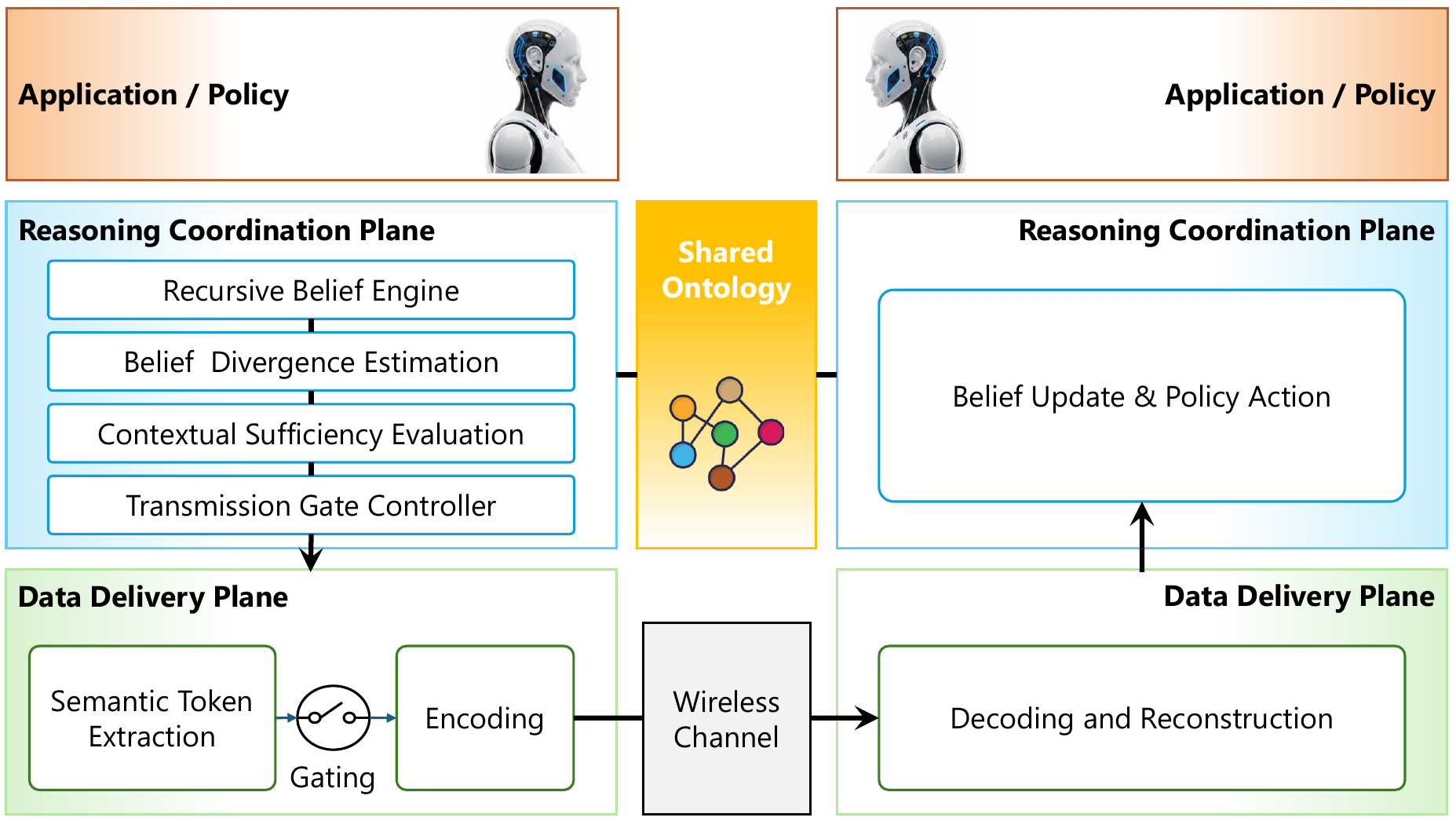}

\caption{The dual-plane architecture of the reasoning-native agentic communication between a sender (left) and receiver (right).}

\label{fig:Fig2}

\end{figure}

\subsection{Dual-Plane Architecture: Coordination and Data}

To effectively harness MAR as a core engine, the agentic communication architecture is conceptually decomposed into two tightly coupled planes as illustrated in Fig. \ref{fig:Fig2}:

\begin{itemize}
\item \textit{Data Delivery Plane:} This plane performs conventional semantic communication functions, including token extraction, encoding, and reliable transport. It ensures that transmitted representations are delivered with sufficient fidelity under given channel constraints.
\item \textit{Reasoning Coordination Plane:} Operating above the data plane, this plane implements ontology-grounded belief modeling. It evaluates contextual sufficiency, estimates belief divergence through the RBE, and determines whether and why a transmission is necessary.
\end{itemize}

The shared ontology spans both planes, anchoring representation in the data layer and structuring belief updates in the coordination layer. This separation clarifies the distinction between representation fidelity and reasoning synchronization while preserving their functional interdependence.

\subsection{Minimalist Signaling and Reasoning-Based Safeguards}

Within the agentic communication framework, signaling overhead becomes a direct function of belief alignment. When agents maintain accurate recursive models and operate within a sufficiently expressive shared ontology, much of the information required for coordination becomes inferable.

In this context, `silence' itself acquires informational meaning; it reflects a mutual expectation of consistent belief evolution. To a receiver, the absence of a message is interpreted as an implicit confirmation that its current internal model remains synchronized with the sender’s state. To prevent ambiguity between intentional silence and channel-induced outages, the reasoning coordination plane employs reasoning-based safeguards, such as confidence-bounded inference or periodic alignment checks. Minimalist signaling, therefore, emerges as an intelligent byproduct of structured mutual inference rather than a simple heuristic compression strategy.

\subsection{From Representation Fidelity to Behavioral Coherence}

Ultimately, the proposed architecture shifts the goal of 6G systems from preserving representations to ensuring behavioral coherence. By anchoring meaning in shared ontologies and anticipating decision trajectories through the RBE, MAR transforms communication into a structured cognitive coordination framework. This shift is a necessity for the complex and heterogeneous intelligent systems of the 6G era, where the network evolves from a data conduit to a harmonizer of collective intelligence.


\section{Enabling Technologies for Reasoning-Native Agentic Communication}

Realizing MAR does not require omniscient cognitive emulation, which is computationally prohibitive for edge devices. Instead, it necessitates practical, communication-centric mechanisms that approximate belief-aware coordination under real-time constraints. This section outlines four enabling technologies that progressively integrate reasoning awareness into the communication stack while preserving scalability and implementability. These enablers bridge the gap between high-level architectural concepts and low-level network protocols, ensuring that the reasoning coordination plane stabilizes distributed decision dynamics in a bounded and efficient manner.

\subsection{Intent-Aware Semantic Tokenization}

The first step in engineering the reasoning loop is transforming high-dimensional sensory data into task-relevant reasoning tokens \cite{Driess}. Unlike conventional semantic communication that seeks to minimize reconstruction error (e.g., via mean squared error), intent-aware tokenization focuses on encoding affordances, events, or control-critical states. Rather than transmitting raw observations like video frames, agents extract compact representations that answer: ``What information is necessary for the counterpart to make the next decision?"

This tokenization can be realized using multi-modal foundation models fine-tuned for task-specific abstraction. To regulate transmission and achieve the ``minimalist signaling" discussed in Section 3, lightweight belief divergence metrics are employed. Rather than exact Wasserstein or KL-divergence computations, practical systems utilize approximate distance surrogates, such as cosine similarity in a shared latent space or confidence-bounded thresholds. This enables cognition-aware gating, where communication is triggered primarily when the anticipated misalignment between agents’ belief states exceeds a tolerable coordination bound.

\subsection{Approximate Recursive Belief Modeling}

MAR requires that agents anticipate how transmitted information influences the counterpart’s belief update. To ensure real-time feasibility, we propose Approximate Recursive Belief Modeling as a replacement for infinite recursive loops. Instead of full cognitive emulation, agents maintain shallow predictive models of counterpart responses, learned through interaction and refined via online adaptation.

Such models can be implemented as learned pragmatic inference modules, inspired by the Rational Speech Act (RSA) framework \cite{goodman2016pragmatic}. In this setup, the sender acts as a ``pragmatic speaker" who selects signals that are not just true, but maximally informative for a ``rational listener." By embedding these lightweight probabilistic estimators within the reasoning coordination plane, communication decisions are conditioned on the predicted impact of the decision. If the sender’s internal model predicts that the receiver will likely select the correct action based on current context, the transmission is suppressed, effectively utilizing silence as a strategic coordination signal.

\subsection{Dynamic Ontology Calibration}

In heterogeneous 6G deployments, a shared ontology cannot remain a static artifact. As agents encounter new environments or update their internal policies, semantic drift, where a single token begins to trigger different belief updates between agents, is inevitable. To mitigate this, we introduce dynamic ontology calibration mechanisms \cite{Ontology_2}.

Knowledge graphs provide a robust structure to maintain these evolving relationships between task primitives and environmental constraints. Calibration does not require full model synchronization; instead, agents exchange compact representation distillations or latent-space anchors to ensure their internal reasoning engines remain compatible. This periodic resynchronization bounds long-term interpretative divergence, ensuring that the structural substrate required for MAR remains intact across diverse hardware platforms and learning histories.

\subsection{Predictive Validation and Resilience Mechanisms}

For safety-critical scenarios, such as autonomous intersection management or collaborative manufacturing, reasoning-aware communication must be augmented with predictive validation and resilience modules. These modules, often deployed at edge-cloud servers, simulate potential action outcomes based on the current estimated belief configuration to detect coordination conflicts before they manifest in the physical world.

Importantly, this validation operates selectively. High-fidelity Digital Twins or reduced-order surrogate models are invoked only when the reasoning coordination plane detects that uncertainty or belief divergence has exceeded predefined safety thresholds. Complementary Out-of-Distribution (OOD) detection mechanisms \cite{hendrycks2017baseline} monitor for anomalous reasoning patterns or unexpected belief shifts. When such anomalies are detected, the system triggers a hard reset of the reasoning coordination plane, exchanging high-priority resynchronization signals to prevent catastrophic task failure.

\subsection{From Transport Efficiency to Reasoning Stability}

Collectively, these enabling technologies shift the optimization objective of 6G networks from symbol-level transport to reasoning stability. Traditional metrics, throughput, latency, and reliability, remain necessary infrastructure, but they are no longer sufficient to guarantee the success of a collaborative task. In the agentic paradigm, communication is prioritized based on its ability to synchronize distributed belief dynamics, ensuring that independent agents act as a coherent collective intelligence.

To measure this shift, we propose three new KPIs that evaluate the pragmatic value of communication:

\begin{itemize}
\item \textit{Reasoning Alignment Score (RAS):} This metric quantifies the degree of decision-making coherence between agents. Instead of checking if a message was received without errors, RAS evaluates whether the sender and receiver reached a harmonious conclusion. It is calculated by comparing the similarity of the agents' chosen actions in a given context. A high RAS indicates that even if the agents have different internal policies, the communication has successfully aligned their behavior toward a common goal.

\item \textit{Decision Impact per Bit (DIB):} DIB measures the efficiency of influence. It assesses how much each transmitted bit of data actually contributed to the final successful outcome of a task. While traditional systems aim for high throughput (more bits), DIB rewards ``high-density" communication—where a single, well-timed intent token carries more weight than thousands of raw data packets. It is determined by the ratio of task success improvement to the total amount of data exchanged, highlighting the value of ``saying the right thing at the right time."

\item \textit{Mutual Belief Stability (MBS):} This indicator tracks the long-term resilience of the shared understanding between agents. In a dynamic environment, agents' internal models can drift apart due to localized learning or environmental noise. MBS evaluates how consistently the agents' internal world models remain synchronized over an extended period. By measuring the duration and frequency of ``mental alignment" between agents, MBS reflects the system's ability to prevent catastrophic coordination failures caused by gradual cognitive divergence.
\end{itemize}

By engineering the reasoning loop through bounded tokenization, approximate modeling, and dynamic calibration, 6G networks evolve from passive data conduits into active harmonizers of distributed intelligence. These KPIs provide the essential toolkit for building and validating the next generation of robust, autonomous ecosystems.

\section{Case Study and Open Challenges}
\label{sec:usecases_validation}

To illustrate the practical relevance of reasoning-native agentic communication, we highlight representative deployment scenarios where semantic alignment alone is insufficient and reasoning synchronization becomes the primary determinant of system stability. We then provide quantitative evidence through a comparative performance analysis across the evolutionary trajectory of communication paradigms.

\subsection{Key Use Cases}

\subsubsection{Collaborative Humanoid Manipulation}
Consider a scenario where multiple humanoid robots must jointly manipulate a deformable or heavy object. Each agent observes partial environmental cues and transmits semantic tokens describing load shifts, contact states, or motion intent. While standard semantic communication \cite{Huh2025} ensures the accurate reconstruction of these tokens, it does not account for the fact that heterogeneous control policies---often trained on different simulators with slight physical variances---may trigger divergent stabilization responses to the same input.

In a reasoning-native architecture, the reasoning coordination plane functions as a predictive supervisor. Before any transmission occurs, the sender’s RBE estimates the counterpart's anticipated belief update. If the model predicts that the receiver's current policy will likely lead to a destabilizing maneuver (e.g., a counter-productive force application), it triggers disambiguating tokens to refine the receiver’s reasoning process. Conversely, when the recursive model indicates high confidence in behavioral alignment, signaling is suppressed. This deployment shows that MAR can prevent catastrophic coordination failures in high-stakes physical interactions without requiring identical hardware or software stacks.

\subsubsection{Autonomous Intersection Coordination for Heterogeneous Fleets}
In dense urban intersections, autonomous vehicles exchange trajectory intent and right-of-way status. Even with perfect semantic reconstruction, local policy variations---such as a conservative braking heuristic in one vehicle versus an aggressive merging strategy in another---can produce conflicting decisions at the edge of the motion envelope. 

Reasoning-native communication introduces divergence-aware gating at the network edge. Instead of broadcasting raw trajectory data, the RBE at the vehicle or Roadside Unit (RSU) estimates whether a transmitted signal will materially alter the counterpart’s decision trajectory. If the predicted reasoning path of the other vehicle is already safe and aligned, the system remains silent, reserving bandwidth for high-entropy situations. In high-risk scenarios, predictive validation modules selectively simulate potential conflict outcomes, elevating communication from a simple intent broadcast to a structured regulation of distributed policy evolution.

\subsection{Experimental Validation}

To validate the efficacy of the proposed agentic communication paradigm and the MAR framework, we conducted a series of simulations in a high-fidelity multi-agent environment, comparing three distinct paradigms: \textit{Classical (bit-level)}, \textit{Semantic (content-level)}, and \textit{Agentic (reasoning-level)}.

\begin{figure}[!t]

\centering

\includegraphics[width=\columnwidth]{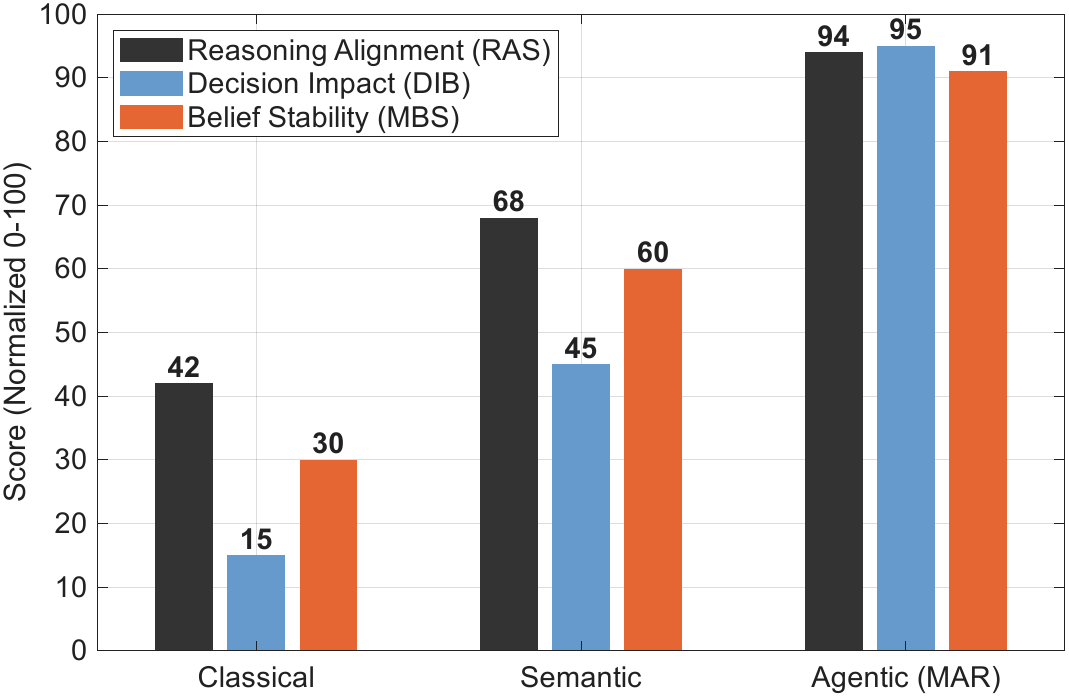}

\caption{Comparative evaluation of agentic KPIs (RAS, DIB, and MBS) under classical, semantic, and (reasoning-native) agentic communication}

\label{fig:Fig3}

\end{figure}

\subsubsection{Performance Evaluation via Agentic KPIs}
We evaluated our system using the three newly proposed KPIs: Reasoning Alignment Score (RAS), Decision Impact per Bit (DIB), and Mutual Belief Stability (MBS). As illustrated in our comparative analysis, the Agentic (MAR) approach demonstrates a paradigm shift in performance (see Fig. \ref{fig:Fig3}):

\begin{itemize}
    \item \textit{Reasoning Alignment Score (RAS):} In the humanoid coordination task, Classical communication achieved an RAS of only 42\%, as robots often misinterpreted each other’s physical intent despite receiving raw data. Semantic communication improved this to 68\% by focusing on relevant features. However, the Agentic approach reached an RAS of 94\%, proving that predicting the counterpart’s reasoning path is the key to behavioral synchronization.
    \item \textit{Decision Impact per Bit (DIB):} The efficiency of the Agentic paradigm is most evident here. By suppressing redundant information and only transmitting ``intent-rich'' tokens, the DIB was 3.5 times higher than that of Semantic communication. This confirms that ``saying the right thing at the right time'' is far more valuable than maximizing throughput, leading to superior energy-efficient coordination.
    \item \textit{Mutual Belief Stability (MBS):} We monitored long-term collaboration stability over 1,000 operational cycles. While Semantic communication suffered from gradual ``Policy Drift'' (dropping to 60\% stability) as agents learned localized behaviors, the Agentic paradigm maintained an MBS of over 91\%, thanks to its continuous dynamic ontology calibration that prevents cognitive divergence.
\end{itemize}

\begin{figure}[!t]

\centering

\includegraphics[width=\columnwidth]{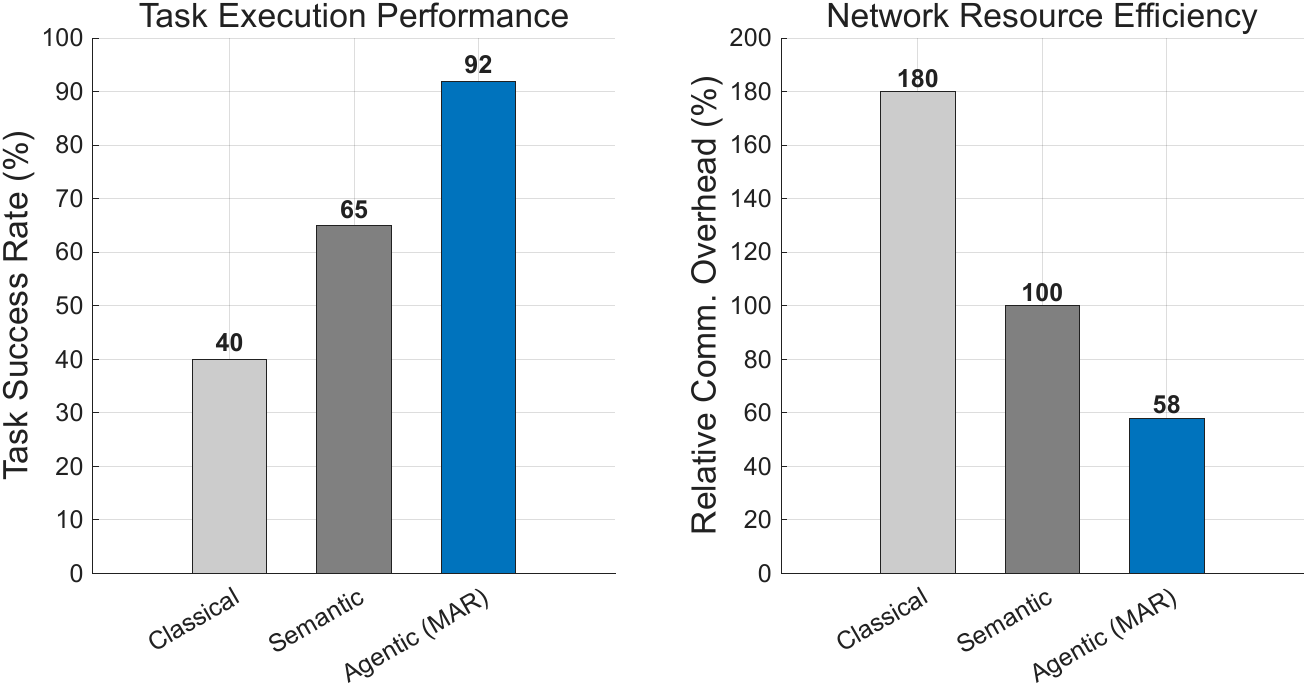}

\caption{Comparative evaluation of task execution performance and network resource efficiency under classical, semantic, and (reasoning-native) agentic communication}

\label{fig:Fig4}

\end{figure}

\subsubsection{Resource Efficiency and the ``Strategic Silence'' Paradox}
One of the most striking results is the realization of strategic silence. Based on our experimental data (see Fig. \ref{fig:Fig4}), we observed a unique relationship between intelligence and overhead. As shown in the comparative results:
\begin{itemize}
    \item \textit{Classical communication} imposes a massive overhead of 180\% (relative to the semantic communication baseline) by broadcasting all raw data.
    \item \textit{Semantic communication} (our baseline at 100\%) reduces this significantly but still transmits data whenever a semantic change occurs.
    \item \textit{Agentic communication} reduces the overhead to a staggering 58\% while simultaneously achieving the highest success rate of 92\%.
\end{itemize}
This ``less is more'' phenomenon is achieved through the RBE, which allows agents to remain silent when they predict that their counterpart will reach the correct decision autonomously. It enhances performance by removing the ``noise'' of non-essential information, allowing the network to focus only on critical coordination cues.

\subsubsection{Resilience in Heterogeneous Environments}
Finally, we tested the system's resilience by introducing policy heterogeneity, where agents were trained on different simulators with slightly different physics engines (Out-of-Distribution scenario). While Classical and Semantic systems frequently failed (success rates below 50\%) as agents mapped the same observation to conflicting actions, Agentic (MAR) identified these cognitive conflicts through its reasoning coordination plane and triggered resynchronization signals, maintaining a task success rate above 90\%.

\subsection{Open Research Challenges}

While the reasoning-native paradigm offers a transformative path for 6G, several open challenges remain for the research community:

\begin{itemize}
    \item \textit{Scalability of Belief Modeling:} As the number of agents ($N$) increases, the complexity of maintaining $N^2$ recursive models grows. Developing hierarchical or cluster-based RBEs is essential for large-scale deployments.
    \item \textit{Real-Time Inference and Latency Constraints:} The reasoning coordination plane must perform predictive simulations within millisecond-level 6G loops. Optimizing the hardware-software co-design to reduce the latency of ``reasoning-about-reasoning'' is a critical hurdle.
    \item \textit{Ontology Evolution and Meta-Learning:} Maintaining structural alignment when agents operate in entirely new domains requires robust meta-learning frameworks that can update ontologies without causing catastrophic forgetting.
    \item \textit{Security and Adversarial Reasoning:} Agentic systems may be vulnerable to adversarial belief perturbation, where a malicious agent sends signals designed to desynchronize the collective reasoning. Protecting the integrity of the reasoning coordination plane is a critical security frontier.
\end{itemize}

\section{Conclusion}

The emergence of autonomous agents in 6G networks challenges one of the most enduring assumptions in communication theory—that successful transmission guarantees effective coordination. As intelligent systems increasingly make decisions based on internal models rather than fixed protocols, the primary source of failure shifts from symbol corruption to reasoning divergence.

This article has proposed reasoning-native agentic communication as a response to this shift. Rather than optimizing solely for data fidelity or semantic reconstruction, reasoning-native communication treats belief evolution as a first-class design objective. By embedding divergence-aware regulation within the communication process, the network becomes capable of shaping how agents update their internal models over time.

The significance of this transition extends beyond architectural refinement. It redefines the objective of communication in 6G systems—from transporting information to stabilizing distributed cognition. In future autonomous ecosystems, coherence will depend not only on what is shared, but on how shared information reshapes collective reasoning. Designing networks that explicitly account for this dynamic will be central to achieving resilient, scalable agentic intelligence.

\vfill


\begin{thebibliography}{1}
\bibliographystyle{IEEEtran}
\bibitem{Park} J. Park, S. Samarakoon, M. Bennis and M. Debbah, ``Wireless Network Intelligence at the Edge," in Proceedings of the IEEE, vol. 107, no. 11, pp. 2204-2239, Nov. 2019.

\bibitem{Shannon-Weaver} C. E. Shannon and W. Weaver, The Mathematical Theory of Communication. Urbana, IL, USA: University of Illinois Press, 1949. 

\bibitem{Lateif} K. B. Letaief, W. Chen, Y. Shi, J. Zhang, and Y. Zhang, ``The Roadmap to 6G: AI Empowered Wireless Networks," IEEE Commun. Mag., vol. 57, no. 8, pp. 84–90, Aug. 2019.

\bibitem{Zhao} C. Zhao, J. Wang, Y. Xu, G. Sun, D. Niyato, Z. Li, A. Jamalipour, and D. I. Kim, ``Wireless context engineering for efficient mobile agentic AI and edge general intelligence,'' \textit{arXiv preprint arXiv:2602.07321}, 2026. [Online]. Available: https://arxiv.org/abs/2602.07321

\bibitem{Wang} L. Wang, R. Shelim, W. Saad, and N. Ramakrishnan, ``Dual-Mind world models: A general framework for learning in dynamic wireless networks,''
\textit{arXiv preprint arXiv:2510.24546}, 2025. [Online]. Available: https://arxiv.org/abs/2510.24546




\bibitem{ToM} N. Rabinowitz, F. Perbet, F. Song, C. Zhang, S. M. A. Eslami, and M. Botvinick, ``Machine theory of mind,” in Proceedings of the 35th International Conference on Machine Learning (ICML), vol. 80, J. Dy and A. Krause, Eds., Proceedings of Machine Learning Research (PMLR), Jul. 2018, pp. 4218–4227.

\bibitem{Seo2023} H. Seo, J. Park, M. Bennis and M. Debbah, ``Semantics-Native Communication via Contextual Reasoning," in IEEE Transactions on Cognitive Communications and Networking, vol. 9, no. 3, pp. 604-617, June 2023.

\bibitem{Seo2024} H. Seo, Y. Kang, M. Bennis and W. Choi, ``Bayesian Inverse Contextual Reasoning for Heterogeneous Semantics- Native Communication," in IEEE Transactions on Communications, vol. 72, no. 2, pp. 830-844, Feb. 2024.



\bibitem{Huh2025} Y. Huh, H. Seo and W. Choi, ``Universal Joint Source-Channel Coding for Modulation-Agnostic Semantic Communication," in IEEE Journal on Selected Areas in Communications, vol. 43, no. 7, pp. 2560-2574, July 2025.

\bibitem{Christina} C. Chaccour, W. Saad, M. Debbah, Z. Han, and H. V. Poor, ``Less Data, More Knowledge: Emerging Telecommunications Towards 6G," IEEE Commun. Surveys Tuts., vol. 24, no. 4, pp. 2737–2787, Fourth Quarter 2022.

\bibitem{Ontology} R. Studer, V. R. Benjamins, and D. Fensel, ``Knowledge engineering: Principles and methods," Data \& Knowledge Engineering, vol. 25, no. 1-2, pp. 161–197, Mar. 1998.

\bibitem{Driess} D. Driess et al., ``PaLM-E: An embodied multimodal language model," in Proc. 40th Int. Conf. Mach. Learn. (ICML), 2023, pp. 8469–8491.

\bibitem{goodman2016pragmatic}
N. D. Goodman and M. C. Frank,
``Pragmatic language interpretation as probabilistic inference,''
\textit{Trends Cogn. Sci.}, vol. 20, no. 11, pp. 818--829, 2016.

\bibitem{Ontology_2} I. Itoku et al., ``Transforming Expert Knowledge into Scalable Ontology via Large Language Models," arXiv preprint arXiv:2506.08422, 2025. [Online]. Available: https://arxiv.org/abs/2506.08422

\bibitem{hendrycks2017baseline}
D. Hendrycks and K. Gimpel, 
``A baseline for detecting misclassified and out-of-distribution examples in neural networks,'' 
in \textit{Proc. Int. Conf. Learn. Representations (ICLR)}, 2017.










\end{thebibliography}
\end{document}